\documentclass[aps,prl,twocolumn,showpacs,preprintnumbers,amsmath,amssymb]{revtex4-1}
\usepackage[breaklinks=true,colorlinks=true
,urlcolor=blue
,anchorcolor=blue
,citecolor=blue
,filecolor=blue
,linkcolor=blue
,menucolor=blue
,pagecolor=blue
,linktocpage=true
,pdfproducer=medialab
,pdfa=true
]{hyperref}
\usepackage{color}
\usepackage[dvipsnames]{xcolor}
\usepackage{graphicx}% Include figure files
\usepackage{dcolumn}% Align table columns on decimal point
\usepackage{bm}% bold math
\usepackage{bbm}

\usepackage{tikz}
\usetikzlibrary{arrows,arrows.meta,intersections, calc,positioning,decorations.pathreplacing,decorations.pathmorphing,shapes}
\usetikzlibrary{patterns}
\usetikzlibrary{decorations.markings}
\usetikzlibrary{knots}
\def\ep{{\epsilon}}

\def\frac#1#2{{#1\over #2}}

\def\s{\sqrt}

\def\be{\begin{equation}}
\def\ee{\end{equation}}
\def\ba{\begin{eqnarray}}
\def\ea{\end{eqnarray}}
\def\de{\partial}

\def\ti{\tilde}

\def\no{\nonumber \\}

\def\la{\langle}
\def\lb{\rangle}
\def\ep{\epsilon}

\def\vp{\varphi}

\usepackage{braket}

\begin{document}

\title{Pseudo Entropy in dS/CFT and Time-like Entanglement Entropy}
\preprint{YITP-22-121; IPMU22-0052}

\author{Kazuki Doi,$^a$ Jonathan Harper,$^a$ Ali Mollabashi$^a$, Tadashi Takayanagi$^{a,b,c}$ and Yusuke Taki$^a$}

\affiliation{$^a$Center for Gravitational Physics, Yukawa Institute for Theoretical Physics, Kyoto University, \\
Kitashirakawa Oiwakecho, Sakyo-ku, Kyoto 606-8502, Japan}

\affiliation{$^b$Inamori Research Institute for Science, 620 Suiginya-cho, Shimogyo-ku, Kyoto 600-8411, Japan}

\affiliation{$^{c}$Kavli Institute for the Physics and Mathematics of the Universe,\\ University of Tokyo, Kashiwa, Chiba 277-8582, Japan}

%\date{\today}

\begin{abstract}
We study holographic entanglement entropy in dS/CFT and introduce time-like entanglement entropy in CFTs. Both of them take complex values in general and are related with each other via an analytical continuation. We argue that they are correctly understood as pseudo entropy. We find that the imaginary part of pseudo entropy implies an emergence of time in dS/CFT.

\end{abstract}

%\pacs{72.10.-d,73.21.-b,73.50.Fq}
% PACS, the Physics and Astronomy
                             % Classification Scheme.
%\keywords{Suggested keywords}%Use showkeys class option if keyword
                              %display desired
\maketitle

%%%%%%%%%%%%%%%%%%%%%%%%%%%%%%
\section{Introduction} 
%%%%%%%%%%%%%%%%%%%%%%%%%%%

Holography in de Sitter space (dS), so called the dS/CFT correspondence \cite{Strominger:2001pn}, has been much more mysterious than that in anti-de Sitter space (AdS) \cite{Maldacena:1997re}. This is mainly because the dual conformal field theory (CFT) is expected to live on a space-like surface and the time coordinate emerges from a Euclidean CFT. Such CFTs turn out to be non-unitary, being exotic compared with text book examples of CFTs. Limited examples of CFTs dual to de Sitter spaces have been known in four dimensional higher spin gravity \cite{Anninos:2011ui} and in three dimensional Einstein gravity \cite{Hikida:2021ese,Hikida:2022ltr}.
Holography in two dimensional de Sitter space has also been developed \cite{Maldacena:2019cbz,Cotler:2019nbi}. One basic way to see the non-unitary nature of dual CFTs is the absence of space-like geodesics which connect two distinct points on the dS boundary at future infinity. This makes the holographic entanglement entropy \cite{Ryu:2006bv,Ryu:2006ef,Hubeny:2007xt} complex-valued \cite{Narayan:2015vda,Sato:2015tta,Miyaji:2015yva,Narayan:2017xca,Narayan:2020nsc,Hikida:2022ltr}, as it involves time-like geodesics.

In this article, we will argue this complex-valued quantity can be properly understood as pseudo entropy introduced in \cite{Nakata:2021ubr}
(refer to \cite{Murciano:2021dga} for a closely related quantity), rather than the standard entanglement entropy, which is real and non-negative. Pseudo entropy is defined as follows. Decomposing the total Hilbert space into those of subsystems $A$ and $B$, we introduce the reduced transition matrix for two pure states $|\psi\lb$ and $|\vp\lb$, by
\ba
\tau_A=\mbox{Tr}_B\left[\frac{|\psi\lb \la \vp|}{\la \vp|\psi\lb}\right].
\ea 
Finally, pseudo entropy is defined by
\ba
S_A=-\mbox{Tr}[\tau_A\log\tau_A].\label{PEdef}
\ea
See \cite{Mollabashi:2020yie,Camilo:2021dtt,Mollabashi:2021xsd,Nishioka:2021cxe,Goto:2021kln,Miyaji:2021lcq,
Akal:2021dqt,Berkooz:2022fso,Akal:2022qei,Mori:2022xec,Mukherjee:2022jac,Guo:2022sfl,Ishiyama:2022odv,
Miyaji:2022cma,Bhattacharya:2022wlp,Guo:2022jzs} for further developments.

The main reason to consider pseudo entropy in dS/CFT is that reduced density matrices in the dual Euclidean CFT are not hermitian. Later we will also point out that an imaginary part of entanglement entropy, which is properly understood as pseudo entropy, naturally arises when we consider a time-like counterpart of entanglement entropy in standard CFTs. This is defined by rotating a space-like subsystem into time-like one, via an analytical continuation. Indeed we will show that the pseudo entropy in dS and the time-like entanglement entropy in AdS/CFT are directly related. Refer to
\cite{PhysRevLett.54.857,Fitzsimons:2013gga,Olson:2011bq,Cotler:2017anu,Cotler:2018sbu,Lerose:2021sag,Giudice:2021smd} for earlier discussions on temporal extension of quantum entanglement, which are different from ours. When we were writing this paper, we noted
the preprint \cite{Liu:2022ugc}, which also analyzes time-like entanglement entropy.
After this paper appeared in arXiv, we noticed the preprint \cite{Narayan:2022afv}, which has a partial overlap.

%%%%%%%%%%%%%%%%%%%%%%%%%%%%%
\section{Pseudo Entropy in dS/CFT}
\begin{figure}[t]
    \centering
    \includegraphics[width=.4\textwidth]{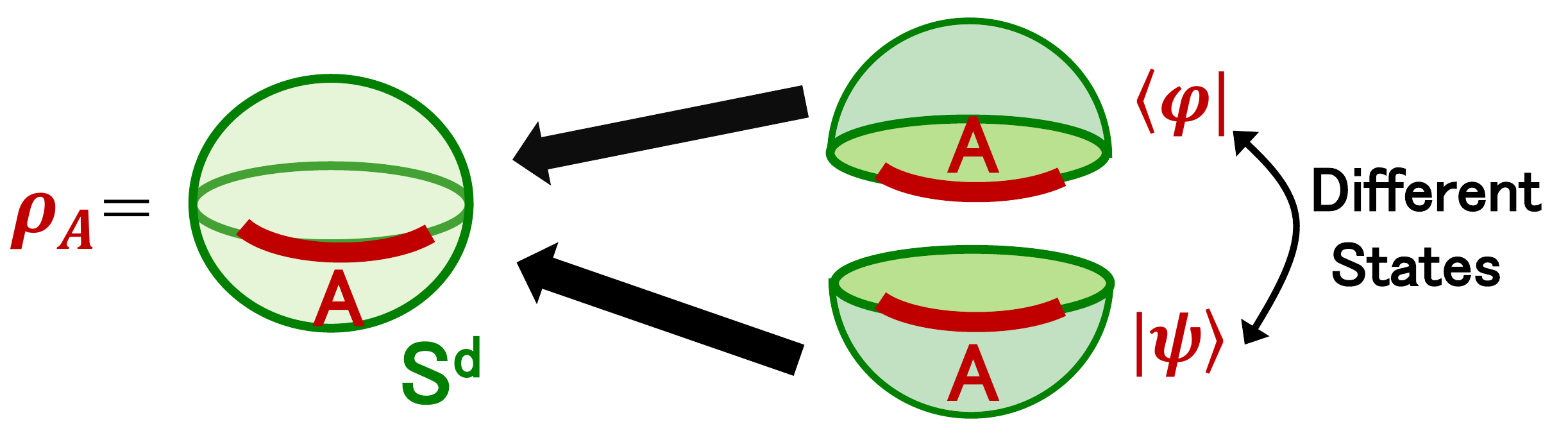}
    \caption{The reduced density matrix $\rho_A$ of a CFT in dS/CFT gets non-hermitian, which leads to pseudo entropy. Here we consider 
    a CFT on $\mathbb{S}^d$, i.e. the boundary of de Sitter space.}
    \label{fig:dSPEstate}
\end{figure}

%%%%%%%%%%%%%%%%%%%%%%%%%%%%%
Consider a $d+1$ dimensional de Sitter space (dS$_{d+1}$), described by a global coordinate
\ba
ds^2=R_{\text{dS}}^2(-d\tau^2+\cosh^2\tau d\Omega_d^2). \label{dSgl}
\ea
We assume the Euclidean instanton i.e. the semi-sphere 
\ba
ds^2=R_{\text{dS}}^2(d\tau_{\text{E}}^2+\cos^2\tau_{\text{E}} d\Omega_d^2),  \label{dSE}
\ea
creates the de Sitter universe at $\tau_{\text{E}}=\tau=0$ and later the Lorentzian evolution occurs 
following (\ref{dSgl}) for $\tau>0$. Then in this setup of the dS/CFT  \cite{Maldacena:2002vr}, the gravity is dual to a Euclidean CFT on $\mathbb{S}^d$.
We can define the reduced density matrix $\rho_A$ by choosing a subsystem $A$ on the equator of $\mathbb{S}^d$ as depicted in Fig. \ref{fig:dSPEstate}.

The dS/CFT relates the CFT partition function $Z_{\text{CFT}}$
to the Hartle-Hawking wave-function of dS \cite{Maldacena:2002vr}: 
\begin{align}
    \Psi_{\text{dS}}[\phi_0]=Z_{\text{CFT}}[\phi_0],
\end{align}
where $\phi_0$ is regarded in the bulk and boundary side as a boundary condition for fields $\phi$ imposed on the future boundary and a source that generates correlation functions, respectively. The wave functional $\Psi_{\text{dS}}$ is given by the path integral over all fields:
\begin{align}
    \Psi_{\text{dS}}[\phi_0]=\int\mathcal{D}\phi\, e^{iS_{\text{G}}[\phi]}\Psi_0,
\end{align}
where $\Psi_0$ denotes the initial state defined by a Euclidean path integral. Therefore $Z_{\text{CFT}}[\phi_0]$ takes complex values as we can easily see from the classical saddle point approximation. For example, assume $d=2$ for simplicity. It is evaluated in terms of the Liouville action \cite{Boruch:2021hqs}:
\ba
&& Z_{\text{CFT}}=e^{-S_{\text{E}}},\no
&& S_{\text{E}}=\frac{ic_{\text{dS}}}{24\pi}\int d^2x \left[(\de_1\phi)^2+(\de_2\phi)^2+\mu e^{2\phi}\right],
\ea
where the metric on $\mathbb{S}^2$ is described by $ds^2=e^{2\phi}(dx_1^2+dx_2^2)$. Here $c_{\text{dS}}$ is defined by 
\begin{align}
    c_{\text{dS}}&=\frac{3R_{\text{dS}}}{2G_{\text{N}}}.
\end{align}
 This means that the path integral on the north and south semi-sphere gives different states and thus the reduced density matrix $\rho_A$ also becomes non-hermitian as illustrated in Fig. \ref{fig:dSPEstate}. In this way, the entanglement entropy in dS/CFT should more properly be regarded as the pseudo entropy. Similar treatment appears in the context of non-hermitian condensed matter systems, see e.g. \cite{Couvreur_2017,Herviou_2019,Chang:2019jcj}.

In the $d=2$ case, if we choose the subsystem $A$ to be an arc with the angle $\theta_0$ on the boundary $\mathbb{S}^2$, the evaluation of the geodesic distance $L_{A}$
between to two boundaries of $A$ leads to 
\begin{align}\label{dSEEgl}
    S_A&=\frac{L_{A}}{4G_{\text{N}}}= -i\frac{c_{\text{dS}}}{3}\log\left[\frac{2}{\ep}\sin\left(\frac{\theta_0}{2}\right)\right]+\frac{\pi c_{\text{dS}}}{6}, 
\end{align}
where the imaginary part comes from the time-like geodesic in (\ref{dSgl}) while the real part does from the space-like one in (\ref{dSE}). We introduced the UV cutoff $\ep$ of the CFT by $e^{\tau_{\infty}}=2/\ep$. This is the holographic pseudo entropy in the global dS$_3$.

In the Poincar\'{e} dS$_3$
\begin{align}
    ds^2=R_{\text{dS}}^2\frac{-d\eta^2+dt^2_{\text{E}}+dx^2}{\eta^2}
\end{align}
the holographic pseudo entropy for an interval $A$ defined by $-x_0/2\leq x\leq x_0/2$ at $t_{\text{E}}=0$ is found to be 
\begin{align}
    S_A=-i\frac{c_{\text{dS}}}{3}\log\left(\frac{x_0}{\epsilon}\right)+\frac{\pi c_{\text{dS}}}{6}.  \label{dspo}
\end{align}
We can obtain these results (\ref{dSEEgl}) and (\ref{dspo}) via the direct computation of geodesic lengths. Equally we can obtain them from the known holographic entanglement entropy via the transformation from the Euclidean AdS (EAdS) to dS:
given by 
\begin{alignat}{3}\label{EAdStodS}
 R_{\text{AdS}}&=-iR_{\text{dS}},\quad & z&=-i\eta,\qquad\quad &(&\text{Poincar\'{e} dS}) \\
 R_{\text{AdS}}&=-iR_{\text{dS}},& \rho&=\tau-\frac{\pi i}{2},&(&\text{Global dS})
\end{alignat}

%%%%%%%%%%%%%%%%%%%%%%%%%%%%%
%%%%%%%%%%%%%%%%%%%%%%%%%%%%%
\section{Time-like entanglement entropy as pseudo entropy }
%%%%%%%%%%%%%%%%%%%%%%%%%%%%%
%%%%%%%%%%%%%%%%%%%%%%%%%%%%%%

Interestingly, when we extend entanglement entropy to time-like subsystems, which we call time-like entanglement entropy, we encounter complex values even for standard unitary CFTs.
The entanglement entropy $S_A$ for an interval $A$ whose time-like and space-like width are given by $T_0$ and $X_0$ reads 
\ba
S_A=\frac{c_{\text{AdS}}}{3}\log\frac{\s{X_0^2-T_0^2}}{\ep},
\ea
where $c_{\text{AdS}}=\frac{3R_{\text{AdS}}}{2G_{\text{N}}}$ is the central charge of the dual CFT \cite{Holzhey:1994we,Calabrese:2004eu}.
The time-like entanglement entropy is obtained by setting $X_0=0$:
\ba
S_A=\frac{c_{\text{AdS}}}{3}\log\left(\frac{T_0}{\ep}\right)+\frac{i\pi c_{\text{AdS}}}{6}.\label{HEET}
\ea
Note that the above procedure can be applied to any unitary CFTs.  We will argue that this quantity is also correctly regarded as pseudo entropy rather than entanglement entropy.

\begin{figure}[h]
    \centering
    \includegraphics[width=.4\textwidth]{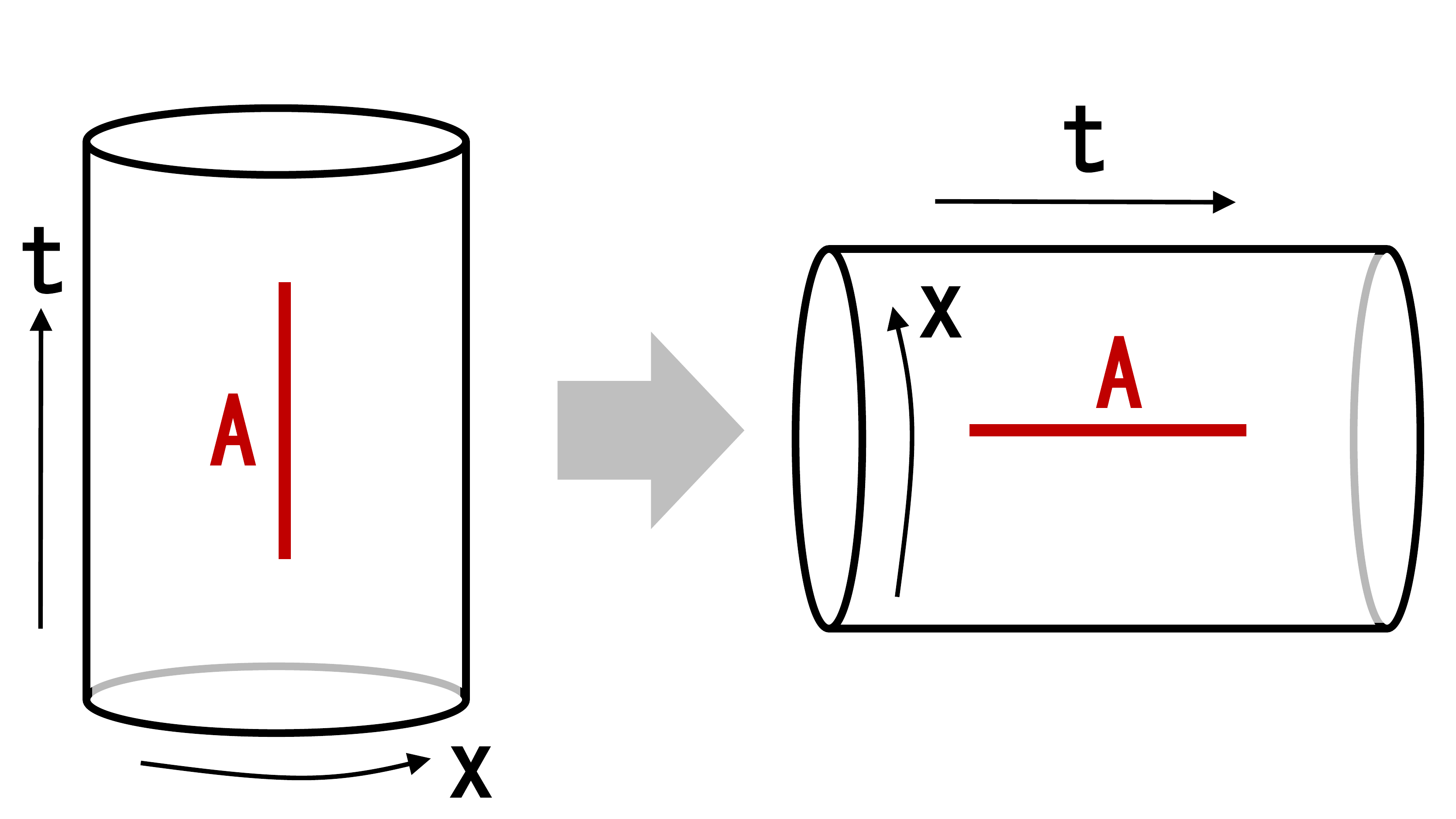}
    \caption{Definition of Time-like Entanglement Entropy}
    \label{fig:QFTPE}
\end{figure}

To see this, we consider its field theoretic calculation. For illustration purpose, consider a free scalar $\Phi$ with a mass $m$ in two dimensions. The space and time coordinate are denoted by 
$x$ and $t$, where the former is compactified with the periodicity $\beta$. The action of the scalar field reads
\ba
S_{\Phi}=\frac{1}{2}\int^\infty_{-\infty} dt \int^{\beta}_0 dx\left[(\de_t\Phi)^2-(\de_x\Phi)^2-m^2\Phi^2\right].
\ea

Now, to calculate the time-like entanglement entropy, we regard $t$ as the ``space" direction and $x$ as the Euclidean time by rotating the spacetime by ninety degree, as depicted in Fig. \ref{fig:QFTPE}.
In this viewpoint we regard the total partition function is written as
\ba
Z_{\Phi}=\mbox{Tr}[e^{-\beta \ti{H}}],
\ea
where the ``Hamiltonian" $\ti{H}$ reads 
\ba
\ti{H}=-\frac{i}{2}\int^{\infty}_{-\infty} dt\left[\Pi^2+(\de_t\Phi)^2-m^2\Phi^2\right].  \label{hamtim}
\ea
Here $\Pi=i\de_x\Phi$ is the canonical momentum. In the massless case $m=0$, in terms of the standard 
positive definite Hamiltonian (we again regard $t$ as a space coordinate)
\ba
H_{\text{CFT}}=\frac{1}{2}\int^{\infty}_{-\infty} dt\left[\Pi^2+(\de_t\Phi)^2\right],
\ea
where the partition function can be rewritten as 
\ba
Z_{\phi}=\mbox{Tr}[e^{i\beta H_{\text{CFT}}}].
\ea
If we trace out the region $B$, the reduced density matrix $\rho_A$ is given by 
\ba
\rho_A=\mbox{Tr}_A[e^{i\beta H_{\text{CFT}}}], \label{reden}
\ea
which is not hermitian. Note also that the time-like entanglement entropy is identical to the entanglement entropy at imaginary temperature. 

We introduce two different states by doubling the Hilbert space similar to the thermofield double:
\ba
&& |\psi\lb=\frac{1}{\s{Z(\delta)}}\sum_n e^{i(\beta+i\delta) E_n/2}|n\lb_1|n\lb_2,\no
&& |\vp\lb=\frac{1}{\s{Z(\delta)}}\sum_n e^{-i(\beta-i\delta) E_n/2}|n\lb_1|n\lb_2,
\ea
such that we obtain 
\ba
\frac{\mbox{Tr}_2|\psi\lb \la\vp|}
{\la\vp|\psi\lb}
=\frac{e^{i(\beta+i\delta)H}}{\mbox{Tr}e^{i(\beta+i\delta)H}},
\ea
where $\delta$ is an infinitesimally small UV regulator. In this way, the time-like entanglement entropy for the reduced density matrix (\ref{reden}) is an example of pseudo entropy (\ref{PEdef}).

\begin{figure}[hhh]
  \centering
  \includegraphics[width=8cm]{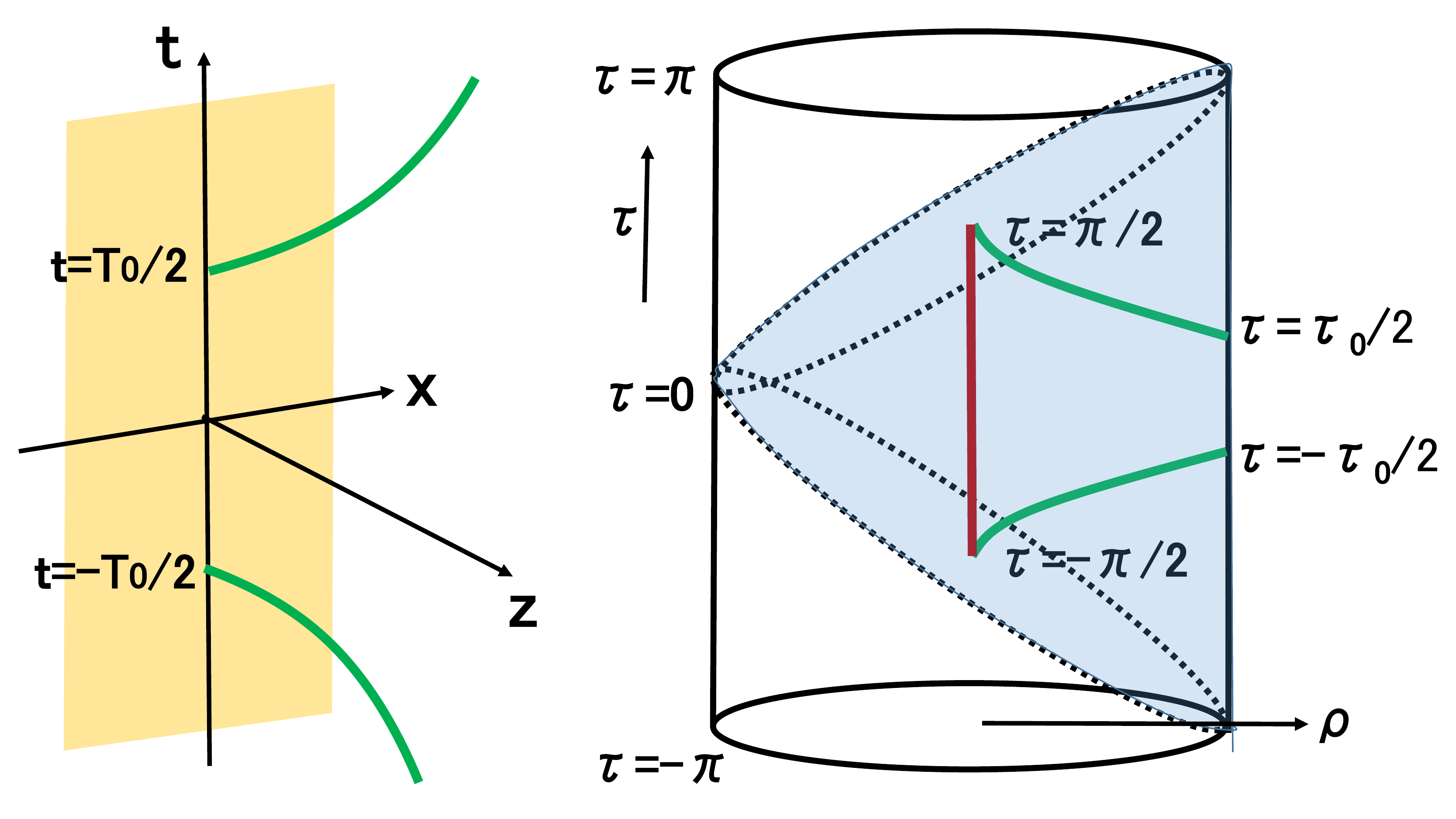}
  \caption{The left panel shows the space-like geodesic (green curve) in the Poincar\'{e} coordinate, which is embedded in the global coordinate with an additional time-like geodesic (red line) in the right panel.} 
\label{fig:HEET}
\end{figure}

In the dual AdS$_3$, we can interpret the time-like entanglement entropy (\ref{HEET}) as a geodesic length as follows. In the Poincar\'{e} coordinate
\be
ds^2=R_{\text{AdS}}^2\frac{dz^2-dt^2+dx^2}{z^2},
\ee
the relevant geodesic is identified with
\ba
t=\s{z^2+T_0^2/4},
\ea
via the Wick rotation of the familiar semi-circle geodesic. 
This is depicted in the left panel of Fig. \ref{fig:HEET}.
Indeed the length of this space-like geodesic reads
\be
S_A=\frac{R_{\text{AdS}}}{4G_{\text{N}}}\cdot 2T_0\int^{\infty}_{\ep} \frac{dz}{z\s{z^2+T_0^2/4}}=\frac{c}{3}\log\frac{T_0}{\ep}.
\ee
This explains the real part of (\ref{HEET}). To understand the imaginary part, we embed the Poincar\'{e} coordinate in the global one:
\be\label{globalc}
ds^2=R_{\text{AdS}}^2(-\cosh^2\rho d\tau^2+d\rho^2+\sinh^2\rho d\theta^2),
\ee
as sketched in the right panel of Fig. \ref{fig:HEET}. The Poincar\'{e} coordinate is covered by the blue region and therefore we need to connect the two endpoints at $\rho=0$ and $\tau=\pm\frac{\pi}{2}$ by a time-like geodesic. 
Since the length is $\pi$, this explains the imaginal part $i\pi\frac{c_{\text{AdS}}}{6}$ of (\ref{HEET}).

If we consider the time-like interval with the length $\tau_0$ in the two dimensional CFT on a cylinder, the above global AdS$_3$ geodesic leads to 
the following estimation of the time-like entanglement entropy
\begin{align}\label{eq:global_TEE}
    S_A=\frac{c_{\text{AdS}}}{3}\log\left[\frac{2}{\ep}\sin\left(\frac{\tau_0}{2}\right)\right]+\frac{c_{\text{AdS}}}{6}\pi i.
\end{align}
It is also useful to note that this can also be obtained by performing the analytical continuation $\beta\to -i\beta$ (remember (\ref{reden}))
on the known finite temperature CFT result for a length $L$ interval $A$ \cite{Calabrese:2004eu}
\ba\label{finiteTEE}
S_A=\frac{c_{\text{AdS}}}{3}\log\left[\frac{\beta}{\pi\ep}\sinh\frac{\pi L}{\beta}\right],
\ea
by setting $\beta=2\pi, L=\tau_0$ with $\ep\to -i\ep$. 

\begin{figure}[h]
    \centering
    \includegraphics[width=.4\textwidth]{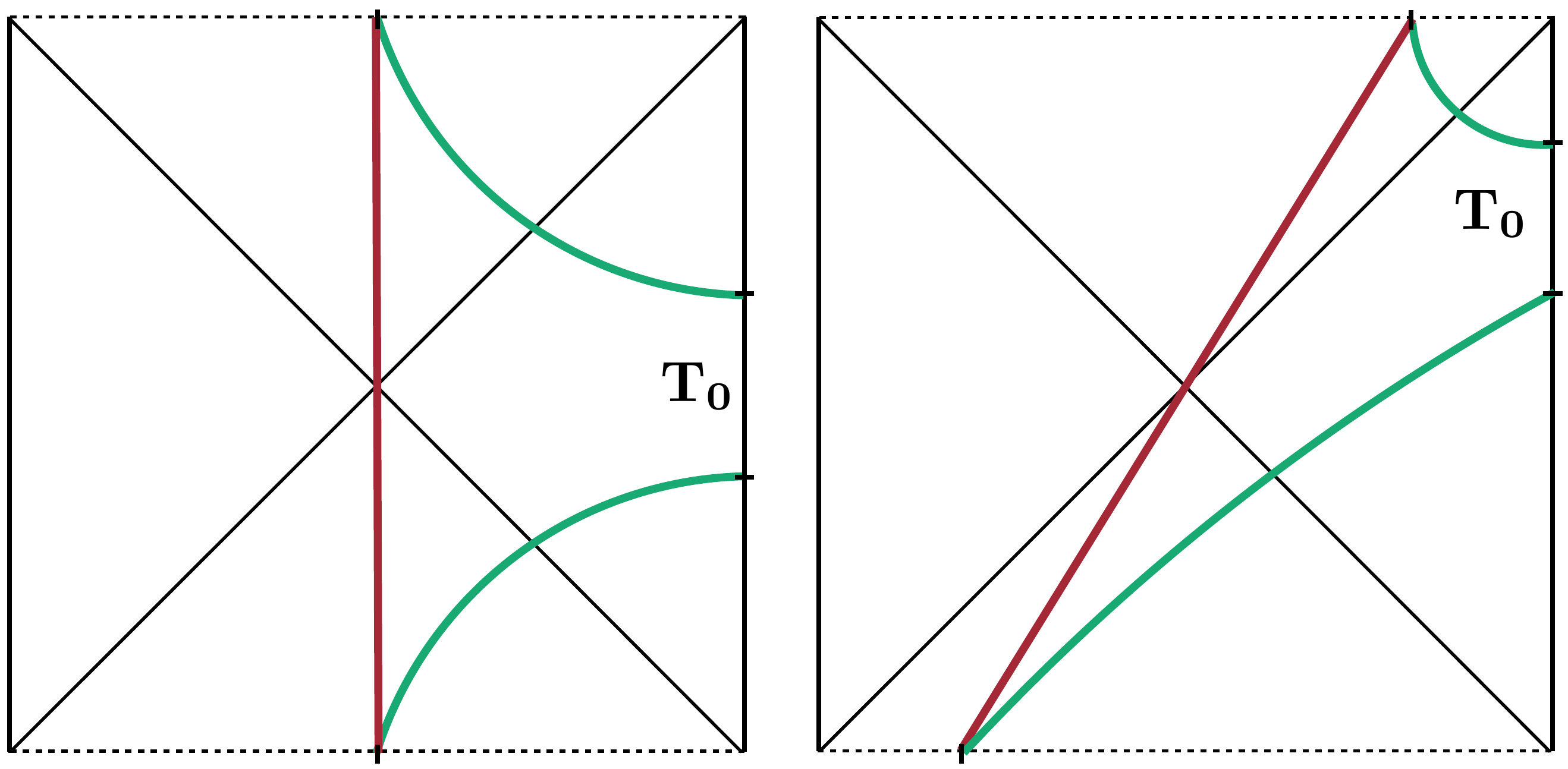}
    \caption{Space and time-like geodesics whose length gives the time-like Holographic Entanglement Entropy in the BTZ geometry. For time intervals not symmetric about the origin (right panel) the space-like geodesics intersect the past and future singularities in different locations.}
    \label{fig:BTZ}
\end{figure}

At finite temperature, the gravity dual is given by the BTZ black hole:
\ba
ds^2=-\left(r^2-r_{+}^2\right)dt^2+\frac{dr^2}{r^2-r_{+}^2}+r^2d\phi^2,
\ea
where $r_+=\frac{2\pi}{\beta}$.
When $A$ is a time-like interval with length $T_0$, the time-like entanglement entropy can be found again from the geodesic length, leading to 
\be
S_A=\frac{c_{\text{AdS}}}{3}\log\left[\frac{\beta}{\pi\ep}\sinh\left(\frac{\pi}{\beta}T_0\right)\right]+\frac{c_{\text{AdS}}}{6}i\pi.
\ee
The space-like and time-like geodesic gives the real and imaginary part as depicted in Fig. \ref{fig:BTZ}.

We would also like to note that the time-like entanglement entropy 
can also be obtained from the de Sitter pseudo entropy by using the relation (\ref{EAdStodS}) between AdS and dS. For example, (\ref{HEET}) is reproduced from (\ref{dspo}) by setting $R_{\text{AdS}}=-iR_{\text{dS}}, z=-i\eta$
and $t=-ix$.

%%%%%%%%%%%%%%%%%%%%%%%%%%%%%%%%%%%%%%
%%%%%%%%%%%%%%%%%%%%%%%%%%%%%%%%%%%%%%
\section{Higher Dimensional Extension}
%%%%%%%%%%%%%%%%%%%%%%%%%%%%%%%%%%%%%%
%%%%%%%%%%%%%%%%%%%%%%%%%%%%%%%%%%%%
We can straightforwardly extend the above holographic calculations to higher dimensions. For simplicity, let us only consider the Poincar\'{e} AdS$_{d+1}$
\begin{align}
    ds^2=R_{\text{AdS}}^2\frac{dz^2-dt^2+dy^2+d\mathbf{x}^2}{z^2},
\end{align}
where $y$ is a direction that we regard as an alternative ``time'' and $\mathbf{x}\in\mathbb{R}^{d-2}$ are the remaining directions. Here we take a hyperbolic subsystem $A$ defined by $t^2-\mathbf{x}^2\ge T_0^2/4$ as a generalization of a temporal interval in the $d=2$ case. Introducing a radial coordinate $\xi=\sqrt{t^2-\mathbf{x}^2}$ for the unit $\mathbb{H}^{d-2}$, the holographic entanglement entropy is evaluated by varying a functional
\begin{align}
    S_A=\frac{R_{\text{AdS}}^{d-1}}{4G_{\text{N}}^{(d+1)}}\text{Vol}(\mathbb{H}^{d-2})\int dz \frac{\xi^{d-2}}{z^{d-1}}\sqrt{1-\xi'(z)^2},
\end{align}
with a boundary condition $\xi(0)=T_0/2$. The resulting extremal surface is the union of a space-like surface
\begin{align}
    \xi^2-z^2=\frac{T_0^2}{4}
\end{align}
and a time-like surface 
\begin{align}
    z^2-\xi^2=\frac{T_0^2}{4}.
\end{align}
This can be regarded as a generalization of Fig. \ref{fig:HEET} in $d=2$. Thus we find 
\begin{align}\label{TEEodd}
    S_A=\frac{R_{\text{AdS}}^{d-1}}{4G_{\text{N}}^{(d+1)}}\text{Vol}(\mathbb{H}^{d-2})&\left[\sum_{k=0}^{\frac{d-3}{2}}\frac{\binom{\frac{d-3}{2}}{k}}{d-2k-2}\left(\frac{T_0}{2\epsilon}\right)^{d-2k-2}\right.\nonumber\\
    &+\left.\frac{i\sqrt{\pi}\Gamma\left(\frac{d-1}{2}\right)}{2\Gamma\left(\frac{d}{2}\right)}\right],
\end{align}
for odd $d$ and 
\begin{align}\label{TEEeven}
    S_A&=\frac{R_{\text{AdS}}^{d-1}}{4G_{\text{N}}^{(d+1)}}\text{Vol}(\mathbb{H}^{d-2})\left[\sum_{k=0}^{\frac{d}{2}-2}\frac{\binom{\frac{d-3}{2}}{k}}{d-2k-2}\left(\frac{T_0}{2\epsilon}\right)^{d-2k-2}\right.\nonumber\\
    &+\frac{\Gamma\left(\frac{d-1}{2}\right)}{\sqrt{\pi}\Gamma\left(\frac{d}{2}\right)}\log\left(\frac{T_0}{2\epsilon}\right)+\left.\frac{i\sqrt{\pi}\Gamma\left(\frac{d-1}{2}\right)}{2\Gamma\left(\frac{d}{2}\right)}\right],
\end{align}
for even $d$, where $\text{Vol}(\mathbb{H}^{d-2})$ denotes the volume of $\mathbb{H}^{d-2}$, assumed to be properly regularized. 

We can relate this result \eqref{TEEodd} to the entanglement entropy for a spherical region \cite{Ryu:2006ef} in EAdS by an analytic continuation $T_0\to -iT_{0},\, \text{Vol}(\mathbb{H}^{d-2})\to i^{d-2}\text{Vol}(\mathbb{S}^{d-2})$. After that, by taking another analytic continuation from EAdS to dS; $R_{\text{AdS}}\to -iR_{\text{dS}},\, \epsilon\to -i\epsilon$, we obtain the pseudo entropy for dS$_{d+1}$/CFT$_d$. It is remarkable that the resulting pseudo entropy has the real part
\begin{align}
    \frac{R_{\text{dS}}^{d-1}\pi^{\frac{d}{2}}}{4G_{\text{N}}^{(d+1)}\Gamma\left(\frac{d}{2}\right)},
\end{align}
which is identical to a half of the de Sitter entropy in dS$_{d+1}$. When $d=2$, this reduces to the real part $\pi c_{\text{dS}}/6$ of \eqref{dSEEgl}. On the other hand, all the divergent terms are purely imaginary, which come from the time-like extremal surfaces in \eqref{dSgl}.

%%%%%%%%%%%%%%%%%%%%%%%%%%%%%%%%%%%%%
%%%%%%%%%%%%%%%%%%%%%%%%%%%%%%%%%%%%%
\section{Numerical Analysis}
\begin{figure}[h]
    \centering
    \includegraphics[width=.4\textwidth]{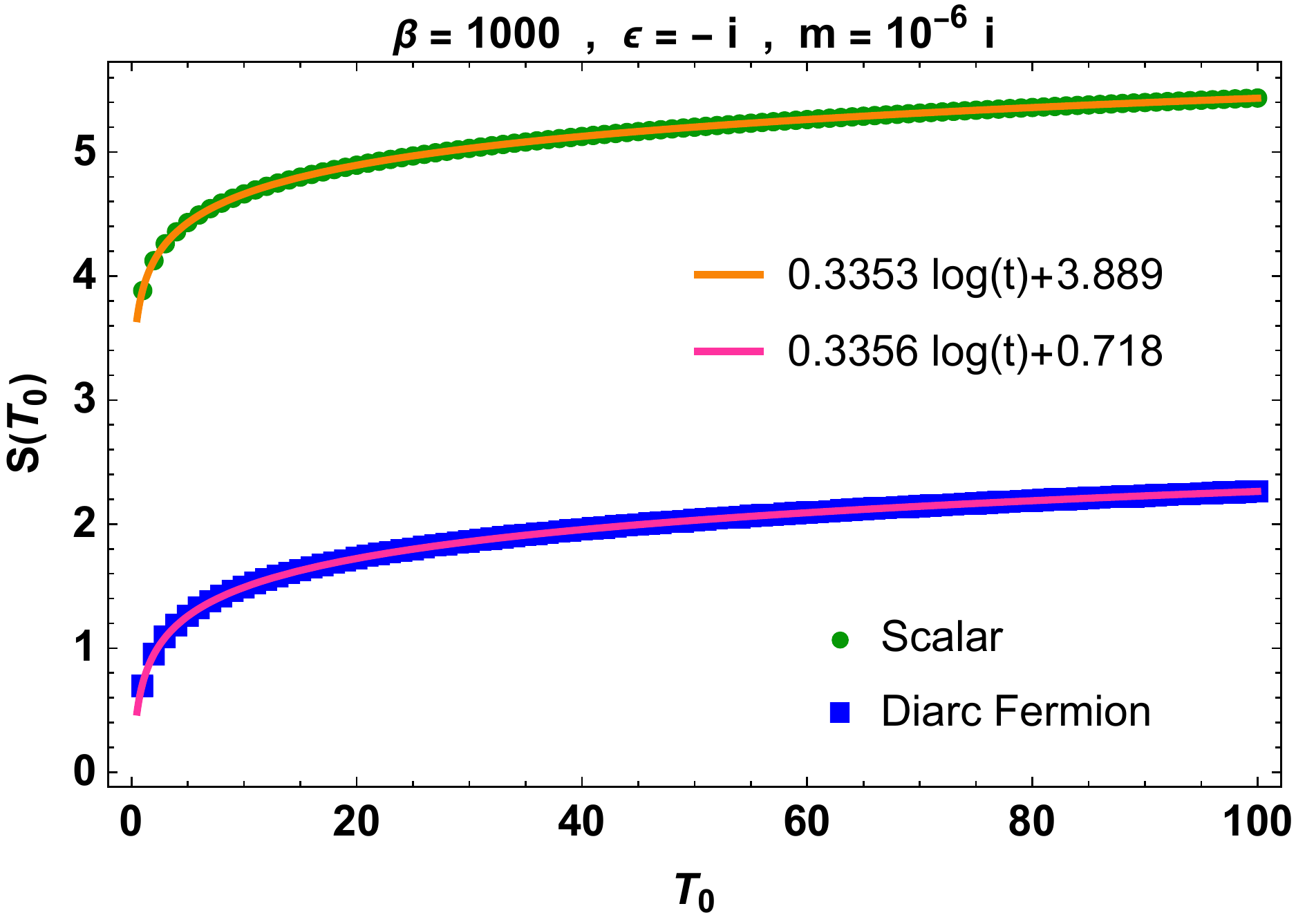}
    \caption{Numerical results for time-like entanglement and the corresponding fit functions for free scalar and Dirac theories.}
    \label{fig:num}
\end{figure}

%%%%%%%%%%%%%%%%%%%%%%%%%%%%%%%%%%%%%%
%%%%%%%%%%%%%%%%%%%%%%%%%%%%%%%%%%%%%
Here we present our numerical checks for the time-like entanglement entropy illustrated in Fig. \ref{fig:QFTPE} for 2d free scalar and free Dirac fermion theories. We adapt the correlator method \cite{Peschel_2003, Latorre:2003kg, Casini:2009sr} to analyse time-like entanglement. For the case of a continuous spatial direction and discrete time direction on an infinite lattice, the relevant correlators for a pure time-like region in the scalar theory are given by
\begin{align}
    \begin{split}
\hspace{-2mm}
\mbox{Tr}[e^{-\beta \ti{H}} \Phi(t)\Phi(t')]
&=\int_{-\pi}^{\pi}\frac{dk}{2\pi}
%\coth\left(\frac{\beta\omega_\Phi(k)}{2}\right)
\frac{\coth\frac{-i\beta\omega_\Phi}{2}}{2\,\omega_\Phi}e^{i k(t-t')}
\\
\hspace{-2mm}
\mbox{Tr}[e^{-\beta \ti{H}} \Pi(t)\Pi(t')]
&=\int_{-\pi}^{\pi}\frac{dk}{2\pi}
\frac{\omega_\Phi\coth\frac{-i\beta\omega_\Phi}{2}}{2}e^{i k(t-t')}
    \end{split}
\end{align}
where $\omega_\Phi(k)=\sqrt{m^2+\frac{4}{\epsilon^2}\sin^2\frac{k}{2}}$ and for the Dirac fermion theory the correlators are given by
\begin{align}
    \begin{split}
\mbox{Tr}[e^{-\beta \ti{H}} \Psi^\dagger(t)\Psi(t')]
&=\frac{\delta_{t,t'}}{2}\mathbf{1}
\\
&
\hspace{-25mm}
-\int_{-\pi}^{\pi}\frac{dk}{2\pi}
\frac{\tanh(-i\beta\omega_\Psi)}{2\,\omega_\Psi}
\begin{pmatrix}
\sin k & m
\\
m & -\sin k
\end{pmatrix}
e^{i k(t-t')}
    \end{split}
\end{align}
where $\omega_\Psi(k)=\sqrt{m^2+\frac{1}{\epsilon^2}\sin^2k}$ (for our conventions see the appendix of \cite{Mozaffar:2021nex}). The important point in both theories is that these correlation functions are the same as the thermal state correlators in a theory with a standard Hamiltonian after applying the analytic continuation $\beta\to -i\beta$. The non-trivial point for our definition of time-like Hamiltonian is that as the expression of $\tilde{H}$ and Fig. \ref{fig:QFTPE} are hinting, we apply $m\to im$ (the IR regulator in the CFT case) and $\epsilon\to -i\epsilon$ (the UV cutoff) in these correlators in order to get time-like entanglement entropy. Fig. \ref{fig:num} shows our numerical results which perfectly agree with our analytic results.

It is worth to note that the imaginary part can be also captured in our numerical method by considering more general regularization prescriptions such as $\epsilon\to -i^a\epsilon$, where $a$ is a real number. The value of $a$ solely affects the coefficient of the imaginary part, which is independent of the subregion length.

%%%%%%%%%%%%%%%%%%%%%%%%%%%%%%%%%%%%%
%%%%%%%%%%%%%%%%%%%%%%%%%%%%%%%%%%%%%
\section{Discussions}
%%%%%%%%%%%%%%%%%%%%%%%%%%%%%%%%%%%%%%
%%%%%%%%%%%%%%%%%%%%%%%%%%%%%%%%%%%%%
In this article, we argued holographic entanglement entropy in dS/CFT and time-like entanglement entropy in ordinary CFTs both should correctly be understood as pseudo entropy. They are related with each other via an analytical continuation. Our results strongly imply the imaginary part of pseudo entropy describes an emergence of time coordinate in holography. This generalizes an emergent space from quantum entanglement \cite{VanRaamsdonk:2010pw,Swingle:2009bg}. We expect this will help us understand the basic mechanism of emergent time in dS/CFT in the near future.

\vspace{5mm}
%%%%%%%%%%%%%%%%%%%%%%%%%%%%%%%%%%%%%
%%%%%%%%%%%%%%%%%%%%%%%%%%%%%
{\bf Acknowledgements} 
%%%%%%%%%%%%%%%%%%%%%%%%%%%%%
%%%%%%%%%%%%%%%%%%%%%%%%%%%%%%%%%%%%%
We are grateful to Yasuaki Hikida, Alexander Jahn and Arthur Parzygnat for useful discussions.
This work is supported by
Grant-in-Aid for Transformative Research Areas (A) No.\,21H05187 and 
JSPS Grant-in-Aid for Scientific Research (A) No.~21H04469.
T.\,T. is supported by the Simons Foundation through the ``It from Qubit'' collaboration, Inamori Research Institute for Science, and
World Premier International Research Center Initiative (WPI Initiative) from the Japan Ministry of Education, Culture, Sports, Science and Technology (MEXT).
Y.\,T. is supported by Grant-in-Aid for JSPS Fellows No.\,22J21950.

\bibliography{dSPE}

%merlin.mbs apsrev4-1.bst 2010-07-25 4.21a (PWD, AO, DPC) hacked
%Control: key (0)
%Control: author (8) initials jnrlst
%Control: editor formatted (1) identically to author
%Control: production of article title (-1) disabled
%Control: page (0) single
%Control: year (1) truncated
%Control: production of eprint (0) enabled
\begin{thebibliography}{55}%
\makeatletter
\providecommand \@ifxundefined [1]{%
 \@ifx{#1\undefined}
}%
\providecommand \@ifnum [1]{%
 \ifnum #1\expandafter \@firstoftwo
 \else \expandafter \@secondoftwo
 \fi
}%
\providecommand \@ifx [1]{%
 \ifx #1\expandafter \@firstoftwo
 \else \expandafter \@secondoftwo
 \fi
}%
\providecommand \natexlab [1]{#1}%
\providecommand \enquote  [1]{``#1''}%
\providecommand \bibnamefont  [1]{#1}%
\providecommand \bibfnamefont [1]{#1}%
\providecommand \citenamefont [1]{#1}%
\providecommand \href@noop [0]{\@secondoftwo}%
\providecommand \href [0]{\begingroup \@sanitize@url \@href}%
\providecommand \@href[1]{\@@startlink{#1}\@@href}%
\providecommand \@@href[1]{\endgroup#1\@@endlink}%
\providecommand \@sanitize@url [0]{\catcode `\\12\catcode `\$12\catcode
  `\&12\catcode `\#12\catcode `\^12\catcode `\_12\catcode `\%12\relax}%
\providecommand \@@startlink[1]{}%
\providecommand \@@endlink[0]{}%
\providecommand \url  [0]{\begingroup\@sanitize@url \@url }%
\providecommand \@url [1]{\endgroup\@href {#1}{\urlprefix }}%
\providecommand \urlprefix  [0]{URL }%
\providecommand \Eprint [0]{\href }%
\providecommand \doibase [0]{http://dx.doi.org/}%
\providecommand \selectlanguage [0]{\@gobble}%
\providecommand \bibinfo  [0]{\@secondoftwo}%
\providecommand \bibfield  [0]{\@secondoftwo}%
\providecommand \translation [1]{[#1]}%
\providecommand \BibitemOpen [0]{}%
\providecommand \bibitemStop [0]{}%
\providecommand \bibitemNoStop [0]{.\EOS\space}%
\providecommand \EOS [0]{\spacefactor3000\relax}%
\providecommand \BibitemShut  [1]{\csname bibitem#1\endcsname}%
\let\auto@bib@innerbib\@empty
%</preamble>
\bibitem [{\citenamefont {Strominger}(2001)}]{Strominger:2001pn}%
  \BibitemOpen
  \bibfield  {author} {\bibinfo {author} {\bibfnamefont {A.}~\bibnamefont
  {Strominger}},\ }\href {\doibase 10.1088/1126-6708/2001/10/034} {\bibfield
  {journal} {\bibinfo  {journal} {JHEP}\ }\textbf {\bibinfo {volume} {10}},\
  \bibinfo {pages} {034} (\bibinfo {year} {2001})},\ \Eprint
  {http://arxiv.org/abs/hep-th/0106113} {arXiv:hep-th/0106113} \BibitemShut
  {NoStop}%
\bibitem [{\citenamefont {Maldacena}(1998)}]{Maldacena:1997re}%
  \BibitemOpen
  \bibfield  {author} {\bibinfo {author} {\bibfnamefont {J.~M.}\ \bibnamefont
  {Maldacena}},\ }\href {\doibase 10.1023/A:1026654312961} {\bibfield
  {journal} {\bibinfo  {journal} {Adv. Theor. Math. Phys.}\ }\textbf {\bibinfo
  {volume} {2}},\ \bibinfo {pages} {231} (\bibinfo {year} {1998})},\ \Eprint
  {http://arxiv.org/abs/hep-th/9711200} {arXiv:hep-th/9711200} \BibitemShut
  {NoStop}%
\bibitem [{\citenamefont {Anninos}\ \emph {et~al.}(2017)\citenamefont
  {Anninos}, \citenamefont {Hartman},\ and\ \citenamefont
  {Strominger}}]{Anninos:2011ui}%
  \BibitemOpen
  \bibfield  {author} {\bibinfo {author} {\bibfnamefont {D.}~\bibnamefont
  {Anninos}}, \bibinfo {author} {\bibfnamefont {T.}~\bibnamefont {Hartman}}, \
  and\ \bibinfo {author} {\bibfnamefont {A.}~\bibnamefont {Strominger}},\
  }\href {\doibase 10.1088/1361-6382/34/1/015009} {\bibfield  {journal}
  {\bibinfo  {journal} {Class. Quant. Grav.}\ }\textbf {\bibinfo {volume}
  {34}},\ \bibinfo {pages} {015009} (\bibinfo {year} {2017})},\ \Eprint
  {http://arxiv.org/abs/1108.5735} {arXiv:1108.5735 [hep-th]} \BibitemShut
  {NoStop}%
\bibitem [{\citenamefont {Hikida}\ \emph
  {et~al.}(2022{\natexlab{a}})\citenamefont {Hikida}, \citenamefont {Nishioka},
  \citenamefont {Takayanagi},\ and\ \citenamefont {Taki}}]{Hikida:2021ese}%
  \BibitemOpen
  \bibfield  {author} {\bibinfo {author} {\bibfnamefont {Y.}~\bibnamefont
  {Hikida}}, \bibinfo {author} {\bibfnamefont {T.}~\bibnamefont {Nishioka}},
  \bibinfo {author} {\bibfnamefont {T.}~\bibnamefont {Takayanagi}}, \ and\
  \bibinfo {author} {\bibfnamefont {Y.}~\bibnamefont {Taki}},\ }\href {\doibase
  10.1103/PhysRevLett.129.041601} {\bibfield  {journal} {\bibinfo  {journal}
  {Phys. Rev. Lett.}\ }\textbf {\bibinfo {volume} {129}},\ \bibinfo {pages}
  {041601} (\bibinfo {year} {2022}{\natexlab{a}})},\ \Eprint
  {http://arxiv.org/abs/2110.03197} {arXiv:2110.03197 [hep-th]} \BibitemShut
  {NoStop}%
\bibitem [{\citenamefont {Hikida}\ \emph
  {et~al.}(2022{\natexlab{b}})\citenamefont {Hikida}, \citenamefont {Nishioka},
  \citenamefont {Takayanagi},\ and\ \citenamefont {Taki}}]{Hikida:2022ltr}%
  \BibitemOpen
  \bibfield  {author} {\bibinfo {author} {\bibfnamefont {Y.}~\bibnamefont
  {Hikida}}, \bibinfo {author} {\bibfnamefont {T.}~\bibnamefont {Nishioka}},
  \bibinfo {author} {\bibfnamefont {T.}~\bibnamefont {Takayanagi}}, \ and\
  \bibinfo {author} {\bibfnamefont {Y.}~\bibnamefont {Taki}},\ }\href {\doibase
  10.1007/JHEP05(2022)129} {\bibfield  {journal} {\bibinfo  {journal} {JHEP}\
  }\textbf {\bibinfo {volume} {05}},\ \bibinfo {pages} {129} (\bibinfo {year}
  {2022}{\natexlab{b}})},\ \Eprint {http://arxiv.org/abs/2203.02852}
  {arXiv:2203.02852 [hep-th]} \BibitemShut {NoStop}%
\bibitem [{\citenamefont {Maldacena}\ \emph {et~al.}(2021)\citenamefont
  {Maldacena}, \citenamefont {Turiaci},\ and\ \citenamefont
  {Yang}}]{Maldacena:2019cbz}%
  \BibitemOpen
  \bibfield  {author} {\bibinfo {author} {\bibfnamefont {J.}~\bibnamefont
  {Maldacena}}, \bibinfo {author} {\bibfnamefont {G.~J.}\ \bibnamefont
  {Turiaci}}, \ and\ \bibinfo {author} {\bibfnamefont {Z.}~\bibnamefont
  {Yang}},\ }\href {\doibase 10.1007/JHEP01(2021)139} {\bibfield  {journal}
  {\bibinfo  {journal} {JHEP}\ }\textbf {\bibinfo {volume} {01}},\ \bibinfo
  {pages} {139} (\bibinfo {year} {2021})},\ \Eprint
  {http://arxiv.org/abs/1904.01911} {arXiv:1904.01911 [hep-th]} \BibitemShut
  {NoStop}%
\bibitem [{\citenamefont {Cotler}\ \emph {et~al.}(2020)\citenamefont {Cotler},
  \citenamefont {Jensen},\ and\ \citenamefont {Maloney}}]{Cotler:2019nbi}%
  \BibitemOpen
  \bibfield  {author} {\bibinfo {author} {\bibfnamefont {J.}~\bibnamefont
  {Cotler}}, \bibinfo {author} {\bibfnamefont {K.}~\bibnamefont {Jensen}}, \
  and\ \bibinfo {author} {\bibfnamefont {A.}~\bibnamefont {Maloney}},\ }\href
  {\doibase 10.1007/JHEP06(2020)048} {\bibfield  {journal} {\bibinfo  {journal}
  {JHEP}\ }\textbf {\bibinfo {volume} {06}},\ \bibinfo {pages} {048} (\bibinfo
  {year} {2020})},\ \Eprint {http://arxiv.org/abs/1905.03780} {arXiv:1905.03780
  [hep-th]} \BibitemShut {NoStop}%
\bibitem [{\citenamefont {Ryu}\ and\ \citenamefont
  {Takayanagi}(2006{\natexlab{a}})}]{Ryu:2006bv}%
  \BibitemOpen
  \bibfield  {author} {\bibinfo {author} {\bibfnamefont {S.}~\bibnamefont
  {Ryu}}\ and\ \bibinfo {author} {\bibfnamefont {T.}~\bibnamefont
  {Takayanagi}},\ }\href {\doibase 10.1103/PhysRevLett.96.181602} {\bibfield
  {journal} {\bibinfo  {journal} {Phys. Rev. Lett.}\ }\textbf {\bibinfo
  {volume} {96}},\ \bibinfo {pages} {181602} (\bibinfo {year}
  {2006}{\natexlab{a}})},\ \Eprint {http://arxiv.org/abs/hep-th/0603001}
  {arXiv:hep-th/0603001} \BibitemShut {NoStop}%
\bibitem [{\citenamefont {Ryu}\ and\ \citenamefont
  {Takayanagi}(2006{\natexlab{b}})}]{Ryu:2006ef}%
  \BibitemOpen
  \bibfield  {author} {\bibinfo {author} {\bibfnamefont {S.}~\bibnamefont
  {Ryu}}\ and\ \bibinfo {author} {\bibfnamefont {T.}~\bibnamefont
  {Takayanagi}},\ }\href {\doibase 10.1088/1126-6708/2006/08/045} {\bibfield
  {journal} {\bibinfo  {journal} {JHEP}\ }\textbf {\bibinfo {volume} {08}},\
  \bibinfo {pages} {045} (\bibinfo {year} {2006}{\natexlab{b}})},\ \Eprint
  {http://arxiv.org/abs/hep-th/0605073} {arXiv:hep-th/0605073} \BibitemShut
  {NoStop}%
\bibitem [{\citenamefont {Hubeny}\ \emph {et~al.}(2007)\citenamefont {Hubeny},
  \citenamefont {Rangamani},\ and\ \citenamefont {Takayanagi}}]{Hubeny:2007xt}%
  \BibitemOpen
  \bibfield  {author} {\bibinfo {author} {\bibfnamefont {V.~E.}\ \bibnamefont
  {Hubeny}}, \bibinfo {author} {\bibfnamefont {M.}~\bibnamefont {Rangamani}}, \
  and\ \bibinfo {author} {\bibfnamefont {T.}~\bibnamefont {Takayanagi}},\
  }\href {\doibase 10.1088/1126-6708/2007/07/062} {\bibfield  {journal}
  {\bibinfo  {journal} {JHEP}\ }\textbf {\bibinfo {volume} {07}},\ \bibinfo
  {pages} {062} (\bibinfo {year} {2007})},\ \Eprint
  {http://arxiv.org/abs/0705.0016} {arXiv:0705.0016 [hep-th]} \BibitemShut
  {NoStop}%
\bibitem [{\citenamefont {Narayan}(2015)}]{Narayan:2015vda}%
  \BibitemOpen
  \bibfield  {author} {\bibinfo {author} {\bibfnamefont {K.}~\bibnamefont
  {Narayan}},\ }\href {\doibase 10.1103/PhysRevD.91.126011} {\bibfield
  {journal} {\bibinfo  {journal} {Phys. Rev. D}\ }\textbf {\bibinfo {volume}
  {91}},\ \bibinfo {pages} {126011} (\bibinfo {year} {2015})},\ \Eprint
  {http://arxiv.org/abs/1501.03019} {arXiv:1501.03019 [hep-th]} \BibitemShut
  {NoStop}%
\bibitem [{\citenamefont {Sato}(2015)}]{Sato:2015tta}%
  \BibitemOpen
  \bibfield  {author} {\bibinfo {author} {\bibfnamefont {Y.}~\bibnamefont
  {Sato}},\ }\href {\doibase 10.1103/PhysRevD.91.086009} {\bibfield  {journal}
  {\bibinfo  {journal} {Phys. Rev. D}\ }\textbf {\bibinfo {volume} {91}},\
  \bibinfo {pages} {086009} (\bibinfo {year} {2015})},\ \Eprint
  {http://arxiv.org/abs/1501.04903} {arXiv:1501.04903 [hep-th]} \BibitemShut
  {NoStop}%
\bibitem [{\citenamefont {Miyaji}\ and\ \citenamefont
  {Takayanagi}(2015)}]{Miyaji:2015yva}%
  \BibitemOpen
  \bibfield  {author} {\bibinfo {author} {\bibfnamefont {M.}~\bibnamefont
  {Miyaji}}\ and\ \bibinfo {author} {\bibfnamefont {T.}~\bibnamefont
  {Takayanagi}},\ }\href {\doibase 10.1093/ptep/ptv089} {\bibfield  {journal}
  {\bibinfo  {journal} {PTEP}\ }\textbf {\bibinfo {volume} {2015}},\ \bibinfo
  {pages} {073B03} (\bibinfo {year} {2015})},\ \Eprint
  {http://arxiv.org/abs/1503.03542} {arXiv:1503.03542 [hep-th]} \BibitemShut
  {NoStop}%
\bibitem [{\citenamefont {Narayan}(2018)}]{Narayan:2017xca}%
  \BibitemOpen
  \bibfield  {author} {\bibinfo {author} {\bibfnamefont {K.}~\bibnamefont
  {Narayan}},\ }\href {\doibase 10.1016/j.physletb.2018.02.010} {\bibfield
  {journal} {\bibinfo  {journal} {Phys. Lett. B}\ }\textbf {\bibinfo {volume}
  {779}},\ \bibinfo {pages} {214} (\bibinfo {year} {2018})},\ \Eprint
  {http://arxiv.org/abs/1711.01107} {arXiv:1711.01107 [hep-th]} \BibitemShut
  {NoStop}%
\bibitem [{\citenamefont {Narayan}(2020)}]{Narayan:2020nsc}%
  \BibitemOpen
  \bibfield  {author} {\bibinfo {author} {\bibfnamefont {K.}~\bibnamefont
  {Narayan}},\ }\href {\doibase 10.1103/PhysRevD.101.086014} {\bibfield
  {journal} {\bibinfo  {journal} {Phys. Rev. D}\ }\textbf {\bibinfo {volume}
  {101}},\ \bibinfo {pages} {086014} (\bibinfo {year} {2020})},\ \Eprint
  {http://arxiv.org/abs/2002.11950} {arXiv:2002.11950 [hep-th]} \BibitemShut
  {NoStop}%
\bibitem [{\citenamefont {Nakata}\ \emph {et~al.}(2021)\citenamefont {Nakata},
  \citenamefont {Takayanagi}, \citenamefont {Taki}, \citenamefont {Tamaoka},\
  and\ \citenamefont {Wei}}]{Nakata:2021ubr}%
  \BibitemOpen
  \bibfield  {author} {\bibinfo {author} {\bibfnamefont {Y.}~\bibnamefont
  {Nakata}}, \bibinfo {author} {\bibfnamefont {T.}~\bibnamefont {Takayanagi}},
  \bibinfo {author} {\bibfnamefont {Y.}~\bibnamefont {Taki}}, \bibinfo {author}
  {\bibfnamefont {K.}~\bibnamefont {Tamaoka}}, \ and\ \bibinfo {author}
  {\bibfnamefont {Z.}~\bibnamefont {Wei}},\ }\href {\doibase
  10.1103/PhysRevD.103.026005} {\bibfield  {journal} {\bibinfo  {journal}
  {Phys. Rev. D}\ }\textbf {\bibinfo {volume} {103}},\ \bibinfo {pages}
  {026005} (\bibinfo {year} {2021})},\ \Eprint
  {http://arxiv.org/abs/2005.13801} {arXiv:2005.13801 [hep-th]} \BibitemShut
  {NoStop}%
\bibitem [{\citenamefont {Murciano}\ \emph {et~al.}(2022)\citenamefont
  {Murciano}, \citenamefont {Calabrese},\ and\ \citenamefont
  {Konik}}]{Murciano:2021dga}%
  \BibitemOpen
  \bibfield  {author} {\bibinfo {author} {\bibfnamefont {S.}~\bibnamefont
  {Murciano}}, \bibinfo {author} {\bibfnamefont {P.}~\bibnamefont {Calabrese}},
  \ and\ \bibinfo {author} {\bibfnamefont {R.~M.}\ \bibnamefont {Konik}},\
  }\href {\doibase 10.1007/JHEP05(2022)152} {\bibfield  {journal} {\bibinfo
  {journal} {JHEP}\ }\textbf {\bibinfo {volume} {05}},\ \bibinfo {pages} {152}
  (\bibinfo {year} {2022})},\ \Eprint {http://arxiv.org/abs/2112.09000}
  {arXiv:2112.09000 [hep-th]} \BibitemShut {NoStop}%
\bibitem [{\citenamefont {Mollabashi}\ \emph
  {et~al.}(2021{\natexlab{a}})\citenamefont {Mollabashi}, \citenamefont
  {Shiba}, \citenamefont {Takayanagi}, \citenamefont {Tamaoka},\ and\
  \citenamefont {Wei}}]{Mollabashi:2020yie}%
  \BibitemOpen
  \bibfield  {author} {\bibinfo {author} {\bibfnamefont {A.}~\bibnamefont
  {Mollabashi}}, \bibinfo {author} {\bibfnamefont {N.}~\bibnamefont {Shiba}},
  \bibinfo {author} {\bibfnamefont {T.}~\bibnamefont {Takayanagi}}, \bibinfo
  {author} {\bibfnamefont {K.}~\bibnamefont {Tamaoka}}, \ and\ \bibinfo
  {author} {\bibfnamefont {Z.}~\bibnamefont {Wei}},\ }\href {\doibase
  10.1103/PhysRevLett.126.081601} {\bibfield  {journal} {\bibinfo  {journal}
  {Phys. Rev. Lett.}\ }\textbf {\bibinfo {volume} {126}},\ \bibinfo {pages}
  {081601} (\bibinfo {year} {2021}{\natexlab{a}})},\ \Eprint
  {http://arxiv.org/abs/2011.09648} {arXiv:2011.09648 [hep-th]} \BibitemShut
  {NoStop}%
\bibitem [{\citenamefont {Camilo}\ and\ \citenamefont
  {Prudenziati}(2021)}]{Camilo:2021dtt}%
  \BibitemOpen
  \bibfield  {author} {\bibinfo {author} {\bibfnamefont {G.}~\bibnamefont
  {Camilo}}\ and\ \bibinfo {author} {\bibfnamefont {A.}~\bibnamefont
  {Prudenziati}},\ }\href@noop {} {\  (\bibinfo {year} {2021})},\ \Eprint
  {http://arxiv.org/abs/2101.02093} {arXiv:2101.02093 [hep-th]} \BibitemShut
  {NoStop}%
\bibitem [{\citenamefont {Mollabashi}\ \emph
  {et~al.}(2021{\natexlab{b}})\citenamefont {Mollabashi}, \citenamefont
  {Shiba}, \citenamefont {Takayanagi}, \citenamefont {Tamaoka},\ and\
  \citenamefont {Wei}}]{Mollabashi:2021xsd}%
  \BibitemOpen
  \bibfield  {author} {\bibinfo {author} {\bibfnamefont {A.}~\bibnamefont
  {Mollabashi}}, \bibinfo {author} {\bibfnamefont {N.}~\bibnamefont {Shiba}},
  \bibinfo {author} {\bibfnamefont {T.}~\bibnamefont {Takayanagi}}, \bibinfo
  {author} {\bibfnamefont {K.}~\bibnamefont {Tamaoka}}, \ and\ \bibinfo
  {author} {\bibfnamefont {Z.}~\bibnamefont {Wei}},\ }\href@noop {} {\
  (\bibinfo {year} {2021}{\natexlab{b}})},\ \Eprint
  {http://arxiv.org/abs/2106.03118} {arXiv:2106.03118 [hep-th]} \BibitemShut
  {NoStop}%
\bibitem [{\citenamefont {Nishioka}\ \emph {et~al.}(2021)\citenamefont
  {Nishioka}, \citenamefont {Takayanagi},\ and\ \citenamefont
  {Taki}}]{Nishioka:2021cxe}%
  \BibitemOpen
  \bibfield  {author} {\bibinfo {author} {\bibfnamefont {T.}~\bibnamefont
  {Nishioka}}, \bibinfo {author} {\bibfnamefont {T.}~\bibnamefont
  {Takayanagi}}, \ and\ \bibinfo {author} {\bibfnamefont {Y.}~\bibnamefont
  {Taki}},\ }\href@noop {} {\  (\bibinfo {year} {2021})},\ \Eprint
  {http://arxiv.org/abs/2107.01797} {arXiv:2107.01797 [hep-th]} \BibitemShut
  {NoStop}%
\bibitem [{\citenamefont {Goto}\ \emph {et~al.}(2021)\citenamefont {Goto},
  \citenamefont {Nozaki},\ and\ \citenamefont {Tamaoka}}]{Goto:2021kln}%
  \BibitemOpen
  \bibfield  {author} {\bibinfo {author} {\bibfnamefont {K.}~\bibnamefont
  {Goto}}, \bibinfo {author} {\bibfnamefont {M.}~\bibnamefont {Nozaki}}, \ and\
  \bibinfo {author} {\bibfnamefont {K.}~\bibnamefont {Tamaoka}},\ }\href
  {\doibase 10.1103/PhysRevD.104.L121902} {\bibfield  {journal} {\bibinfo
  {journal} {Phys. Rev. D}\ }\textbf {\bibinfo {volume} {104}},\ \bibinfo
  {pages} {L121902} (\bibinfo {year} {2021})},\ \Eprint
  {http://arxiv.org/abs/2109.00372} {arXiv:2109.00372 [hep-th]} \BibitemShut
  {NoStop}%
\bibitem [{\citenamefont {Miyaji}(2021)}]{Miyaji:2021lcq}%
  \BibitemOpen
  \bibfield  {author} {\bibinfo {author} {\bibfnamefont {M.}~\bibnamefont
  {Miyaji}},\ }\href {\doibase 10.1007/JHEP12(2021)013} {\bibfield  {journal}
  {\bibinfo  {journal} {JHEP}\ }\textbf {\bibinfo {volume} {12}},\ \bibinfo
  {pages} {013} (\bibinfo {year} {2021})},\ \Eprint
  {http://arxiv.org/abs/2109.03830} {arXiv:2109.03830 [hep-th]} \BibitemShut
  {NoStop}%
\bibitem [{\citenamefont {Akal}\ \emph
  {et~al.}(2022{\natexlab{a}})\citenamefont {Akal}, \citenamefont {Kawamoto},
  \citenamefont {Ruan}, \citenamefont {Takayanagi},\ and\ \citenamefont
  {Wei}}]{Akal:2021dqt}%
  \BibitemOpen
  \bibfield  {author} {\bibinfo {author} {\bibfnamefont {I.}~\bibnamefont
  {Akal}}, \bibinfo {author} {\bibfnamefont {T.}~\bibnamefont {Kawamoto}},
  \bibinfo {author} {\bibfnamefont {S.-M.}\ \bibnamefont {Ruan}}, \bibinfo
  {author} {\bibfnamefont {T.}~\bibnamefont {Takayanagi}}, \ and\ \bibinfo
  {author} {\bibfnamefont {Z.}~\bibnamefont {Wei}},\ }\href {\doibase
  10.1103/PhysRevD.105.126026} {\bibfield  {journal} {\bibinfo  {journal}
  {Phys. Rev. D}\ }\textbf {\bibinfo {volume} {105}},\ \bibinfo {pages}
  {126026} (\bibinfo {year} {2022}{\natexlab{a}})},\ \Eprint
  {http://arxiv.org/abs/2112.08433} {arXiv:2112.08433 [hep-th]} \BibitemShut
  {NoStop}%
\bibitem [{\citenamefont {Berkooz}\ \emph {et~al.}(2022)\citenamefont
  {Berkooz}, \citenamefont {Brukner}, \citenamefont {Ross},\ and\ \citenamefont
  {Watanabe}}]{Berkooz:2022fso}%
  \BibitemOpen
  \bibfield  {author} {\bibinfo {author} {\bibfnamefont {M.}~\bibnamefont
  {Berkooz}}, \bibinfo {author} {\bibfnamefont {N.}~\bibnamefont {Brukner}},
  \bibinfo {author} {\bibfnamefont {S.~F.}\ \bibnamefont {Ross}}, \ and\
  \bibinfo {author} {\bibfnamefont {M.}~\bibnamefont {Watanabe}},\ }\href
  {\doibase 10.1007/JHEP08(2022)051} {\bibfield  {journal} {\bibinfo  {journal}
  {JHEP}\ }\textbf {\bibinfo {volume} {08}},\ \bibinfo {pages} {051} (\bibinfo
  {year} {2022})},\ \Eprint {http://arxiv.org/abs/2202.11381} {arXiv:2202.11381
  [hep-th]} \BibitemShut {NoStop}%
\bibitem [{\citenamefont {Akal}\ \emph
  {et~al.}(2022{\natexlab{b}})\citenamefont {Akal}, \citenamefont {Kawamoto},
  \citenamefont {Ruan}, \citenamefont {Takayanagi},\ and\ \citenamefont
  {Wei}}]{Akal:2022qei}%
  \BibitemOpen
  \bibfield  {author} {\bibinfo {author} {\bibfnamefont {I.}~\bibnamefont
  {Akal}}, \bibinfo {author} {\bibfnamefont {T.}~\bibnamefont {Kawamoto}},
  \bibinfo {author} {\bibfnamefont {S.-M.}\ \bibnamefont {Ruan}}, \bibinfo
  {author} {\bibfnamefont {T.}~\bibnamefont {Takayanagi}}, \ and\ \bibinfo
  {author} {\bibfnamefont {Z.}~\bibnamefont {Wei}},\ }\href {\doibase
  10.1007/JHEP08(2022)296} {\bibfield  {journal} {\bibinfo  {journal} {JHEP}\
  }\textbf {\bibinfo {volume} {08}},\ \bibinfo {pages} {296} (\bibinfo {year}
  {2022}{\natexlab{b}})},\ \Eprint {http://arxiv.org/abs/2205.02663}
  {arXiv:2205.02663 [hep-th]} \BibitemShut {NoStop}%
\bibitem [{\citenamefont {Mori}\ \emph {et~al.}(2022)\citenamefont {Mori},
  \citenamefont {Manabe},\ and\ \citenamefont {Matsueda}}]{Mori:2022xec}%
  \BibitemOpen
  \bibfield  {author} {\bibinfo {author} {\bibfnamefont {T.}~\bibnamefont
  {Mori}}, \bibinfo {author} {\bibfnamefont {H.}~\bibnamefont {Manabe}}, \ and\
  \bibinfo {author} {\bibfnamefont {H.}~\bibnamefont {Matsueda}},\ }\href@noop
  {} {\  (\bibinfo {year} {2022})},\ \Eprint {http://arxiv.org/abs/2205.06633}
  {arXiv:2205.06633 [hep-th]} \BibitemShut {NoStop}%
\bibitem [{\citenamefont {Mukherjee}(2022)}]{Mukherjee:2022jac}%
  \BibitemOpen
  \bibfield  {author} {\bibinfo {author} {\bibfnamefont {J.}~\bibnamefont
  {Mukherjee}},\ }\href {\doibase 10.1007/JHEP10(2022)016} {\bibfield
  {journal} {\bibinfo  {journal} {JHEP}\ }\textbf {\bibinfo {volume} {10}},\
  \bibinfo {pages} {016} (\bibinfo {year} {2022})},\ \Eprint
  {http://arxiv.org/abs/2205.08179} {arXiv:2205.08179 [hep-th]} \BibitemShut
  {NoStop}%
\bibitem [{\citenamefont {Guo}\ \emph {et~al.}(2022{\natexlab{a}})\citenamefont
  {Guo}, \citenamefont {He},\ and\ \citenamefont {Zhang}}]{Guo:2022sfl}%
  \BibitemOpen
  \bibfield  {author} {\bibinfo {author} {\bibfnamefont {W.-z.}\ \bibnamefont
  {Guo}}, \bibinfo {author} {\bibfnamefont {S.}~\bibnamefont {He}}, \ and\
  \bibinfo {author} {\bibfnamefont {Y.-X.}\ \bibnamefont {Zhang}},\ }\href
  {\doibase 10.1007/JHEP09(2022)094} {\bibfield  {journal} {\bibinfo  {journal}
  {JHEP}\ }\textbf {\bibinfo {volume} {09}},\ \bibinfo {pages} {094} (\bibinfo
  {year} {2022}{\natexlab{a}})},\ \Eprint {http://arxiv.org/abs/2206.11818}
  {arXiv:2206.11818 [hep-th]} \BibitemShut {NoStop}%
\bibitem [{\citenamefont {Ishiyama}\ \emph {et~al.}(2022)\citenamefont
  {Ishiyama}, \citenamefont {Kojima}, \citenamefont {Matsui},\ and\
  \citenamefont {Tamaoka}}]{Ishiyama:2022odv}%
  \BibitemOpen
  \bibfield  {author} {\bibinfo {author} {\bibfnamefont {Y.}~\bibnamefont
  {Ishiyama}}, \bibinfo {author} {\bibfnamefont {R.}~\bibnamefont {Kojima}},
  \bibinfo {author} {\bibfnamefont {S.}~\bibnamefont {Matsui}}, \ and\ \bibinfo
  {author} {\bibfnamefont {K.}~\bibnamefont {Tamaoka}},\ }\href@noop {} {\
  (\bibinfo {year} {2022})},\ \Eprint {http://arxiv.org/abs/2206.14551}
  {arXiv:2206.14551 [hep-th]} \BibitemShut {NoStop}%
\bibitem [{\citenamefont {Miyaji}\ and\ \citenamefont
  {Murdia}(2022)}]{Miyaji:2022cma}%
  \BibitemOpen
  \bibfield  {author} {\bibinfo {author} {\bibfnamefont {M.}~\bibnamefont
  {Miyaji}}\ and\ \bibinfo {author} {\bibfnamefont {C.}~\bibnamefont
  {Murdia}},\ }\href@noop {} {\  (\bibinfo {year} {2022})},\ \Eprint
  {http://arxiv.org/abs/2208.13783} {arXiv:2208.13783 [hep-th]} \BibitemShut
  {NoStop}%
\bibitem [{\citenamefont {Bhattacharya}\ \emph {et~al.}(2022)\citenamefont
  {Bhattacharya}, \citenamefont {Bhattacharyya},\ and\ \citenamefont
  {Maulik}}]{Bhattacharya:2022wlp}%
  \BibitemOpen
  \bibfield  {author} {\bibinfo {author} {\bibfnamefont {A.}~\bibnamefont
  {Bhattacharya}}, \bibinfo {author} {\bibfnamefont {A.}~\bibnamefont
  {Bhattacharyya}}, \ and\ \bibinfo {author} {\bibfnamefont {S.}~\bibnamefont
  {Maulik}},\ }\href@noop {} {\  (\bibinfo {year} {2022})},\ \Eprint
  {http://arxiv.org/abs/2209.00049} {arXiv:2209.00049 [hep-th]} \BibitemShut
  {NoStop}%
\bibitem [{\citenamefont {Guo}\ \emph {et~al.}(2022{\natexlab{b}})\citenamefont
  {Guo}, \citenamefont {He},\ and\ \citenamefont {Zhang}}]{Guo:2022jzs}%
  \BibitemOpen
  \bibfield  {author} {\bibinfo {author} {\bibfnamefont {W.-z.}\ \bibnamefont
  {Guo}}, \bibinfo {author} {\bibfnamefont {S.}~\bibnamefont {He}}, \ and\
  \bibinfo {author} {\bibfnamefont {Y.-X.}\ \bibnamefont {Zhang}},\ }\href@noop
  {} {\  (\bibinfo {year} {2022}{\natexlab{b}})},\ \Eprint
  {http://arxiv.org/abs/2209.07308} {arXiv:2209.07308 [hep-th]} \BibitemShut
  {NoStop}%
\bibitem [{\citenamefont {Leggett}\ and\ \citenamefont
  {Garg}(1985)}]{PhysRevLett.54.857}%
  \BibitemOpen
  \bibfield  {author} {\bibinfo {author} {\bibfnamefont {A.~J.}\ \bibnamefont
  {Leggett}}\ and\ \bibinfo {author} {\bibfnamefont {A.}~\bibnamefont {Garg}},\
  }\href {\doibase 10.1103/PhysRevLett.54.857} {\bibfield  {journal} {\bibinfo
  {journal} {Phys. Rev. Lett.}\ }\textbf {\bibinfo {volume} {54}},\ \bibinfo
  {pages} {857} (\bibinfo {year} {1985})}\BibitemShut {NoStop}%
\bibitem [{\citenamefont {Fitzsimons}\ \emph {et~al.}(2013)\citenamefont
  {Fitzsimons}, \citenamefont {Jones},\ and\ \citenamefont
  {Vedral}}]{Fitzsimons:2013gga}%
  \BibitemOpen
  \bibfield  {author} {\bibinfo {author} {\bibfnamefont {J.}~\bibnamefont
  {Fitzsimons}}, \bibinfo {author} {\bibfnamefont {J.}~\bibnamefont {Jones}}, \
  and\ \bibinfo {author} {\bibfnamefont {V.}~\bibnamefont {Vedral}},\
  }\href@noop {} {\  (\bibinfo {year} {2013})},\ \Eprint
  {http://arxiv.org/abs/1302.2731} {arXiv:1302.2731 [quant-ph]} \BibitemShut
  {NoStop}%
\bibitem [{\citenamefont {Olson}\ and\ \citenamefont
  {Ralph}(2012)}]{Olson:2011bq}%
  \BibitemOpen
  \bibfield  {author} {\bibinfo {author} {\bibfnamefont {S.~J.}\ \bibnamefont
  {Olson}}\ and\ \bibinfo {author} {\bibfnamefont {T.~C.}\ \bibnamefont
  {Ralph}},\ }\href {\doibase 10.1103/PhysRevA.85.012306} {\bibfield  {journal}
  {\bibinfo  {journal} {Phys. Rev. A}\ }\textbf {\bibinfo {volume} {85}},\
  \bibinfo {pages} {012306} (\bibinfo {year} {2012})},\ \Eprint
  {http://arxiv.org/abs/1101.2565} {arXiv:1101.2565 [quant-ph]} \BibitemShut
  {NoStop}%
\bibitem [{\citenamefont {Cotler}\ \emph {et~al.}(2018)\citenamefont {Cotler},
  \citenamefont {Jian}, \citenamefont {Qi},\ and\ \citenamefont
  {Wilczek}}]{Cotler:2017anu}%
  \BibitemOpen
  \bibfield  {author} {\bibinfo {author} {\bibfnamefont {J.}~\bibnamefont
  {Cotler}}, \bibinfo {author} {\bibfnamefont {C.-M.}\ \bibnamefont {Jian}},
  \bibinfo {author} {\bibfnamefont {X.-L.}\ \bibnamefont {Qi}}, \ and\ \bibinfo
  {author} {\bibfnamefont {F.}~\bibnamefont {Wilczek}},\ }\href {\doibase
  10.1007/JHEP09(2018)093} {\bibfield  {journal} {\bibinfo  {journal} {JHEP}\
  }\textbf {\bibinfo {volume} {09}},\ \bibinfo {pages} {093} (\bibinfo {year}
  {2018})},\ \Eprint {http://arxiv.org/abs/1711.03119} {arXiv:1711.03119
  [quant-ph]} \BibitemShut {NoStop}%
\bibitem [{\citenamefont {Cotler}\ \emph {et~al.}(2019)\citenamefont {Cotler}
  \emph {et~al.}}]{Cotler:2018sbu}%
  \BibitemOpen
  \bibfield  {author} {\bibinfo {author} {\bibfnamefont {J.}~\bibnamefont
  {Cotler}} \emph {et~al.},\ }\href {\doibase 10.1103/PhysRevX.9.031013}
  {\bibfield  {journal} {\bibinfo  {journal} {Phys. Rev. X}\ }\textbf {\bibinfo
  {volume} {9}},\ \bibinfo {pages} {031013} (\bibinfo {year} {2019})},\ \Eprint
  {http://arxiv.org/abs/1812.02175} {arXiv:1812.02175 [quant-ph]} \BibitemShut
  {NoStop}%
\bibitem [{\citenamefont {Lerose}\ \emph {et~al.}(2021)\citenamefont {Lerose},
  \citenamefont {Sonner},\ and\ \citenamefont {Abanin}}]{Lerose:2021sag}%
  \BibitemOpen
  \bibfield  {author} {\bibinfo {author} {\bibfnamefont {A.}~\bibnamefont
  {Lerose}}, \bibinfo {author} {\bibfnamefont {M.}~\bibnamefont {Sonner}}, \
  and\ \bibinfo {author} {\bibfnamefont {D.~A.}\ \bibnamefont {Abanin}},\
  }\href {\doibase 10.1103/PhysRevB.104.035137} {\bibfield  {journal} {\bibinfo
   {journal} {Phys. Rev. B}\ }\textbf {\bibinfo {volume} {104}},\ \bibinfo
  {pages} {035137} (\bibinfo {year} {2021})},\ \Eprint
  {http://arxiv.org/abs/2104.07607} {arXiv:2104.07607 [quant-ph]} \BibitemShut
  {NoStop}%
\bibitem [{\citenamefont {Giudice}\ \emph {et~al.}(2022)\citenamefont
  {Giudice}, \citenamefont {Giudici}, \citenamefont {Sonner}, \citenamefont
  {Thoenniss}, \citenamefont {Lerose}, \citenamefont {Abanin},\ and\
  \citenamefont {Piroli}}]{Giudice:2021smd}%
  \BibitemOpen
  \bibfield  {author} {\bibinfo {author} {\bibfnamefont {G.}~\bibnamefont
  {Giudice}}, \bibinfo {author} {\bibfnamefont {G.}~\bibnamefont {Giudici}},
  \bibinfo {author} {\bibfnamefont {M.}~\bibnamefont {Sonner}}, \bibinfo
  {author} {\bibfnamefont {J.}~\bibnamefont {Thoenniss}}, \bibinfo {author}
  {\bibfnamefont {A.}~\bibnamefont {Lerose}}, \bibinfo {author} {\bibfnamefont
  {D.~A.}\ \bibnamefont {Abanin}}, \ and\ \bibinfo {author} {\bibfnamefont
  {L.}~\bibnamefont {Piroli}},\ }\href {\doibase
  10.1103/PhysRevLett.128.220401} {\bibfield  {journal} {\bibinfo  {journal}
  {Phys. Rev. Lett.}\ }\textbf {\bibinfo {volume} {128}},\ \bibinfo {pages}
  {220401} (\bibinfo {year} {2022})},\ \Eprint
  {http://arxiv.org/abs/2112.14264} {arXiv:2112.14264 [cond-mat.stat-mech]}
  \BibitemShut {NoStop}%
\bibitem [{\citenamefont {Liu}\ \emph {et~al.}(2022)\citenamefont {Liu},
  \citenamefont {Chen},\ and\ \citenamefont {Lian}}]{Liu:2022ugc}%
  \BibitemOpen
  \bibfield  {author} {\bibinfo {author} {\bibfnamefont {B.}~\bibnamefont
  {Liu}}, \bibinfo {author} {\bibfnamefont {H.}~\bibnamefont {Chen}}, \ and\
  \bibinfo {author} {\bibfnamefont {B.}~\bibnamefont {Lian}},\ }\href@noop {}
  {\  (\bibinfo {year} {2022})},\ \Eprint {http://arxiv.org/abs/2210.03134}
  {arXiv:2210.03134 [cond-mat.stat-mech]} \BibitemShut {NoStop}%
\bibitem [{\citenamefont {Narayan}(2022)}]{Narayan:2022afv}%
  \BibitemOpen
  \bibfield  {author} {\bibinfo {author} {\bibfnamefont {K.}~\bibnamefont
  {Narayan}},\ }\href@noop {} {\  (\bibinfo {year} {2022})},\ \Eprint
  {http://arxiv.org/abs/2210.12963} {arXiv:2210.12963 [hep-th]} \BibitemShut
  {NoStop}%
\bibitem [{\citenamefont {Maldacena}(2003)}]{Maldacena:2002vr}%
  \BibitemOpen
  \bibfield  {author} {\bibinfo {author} {\bibfnamefont {J.~M.}\ \bibnamefont
  {Maldacena}},\ }\href {\doibase 10.1088/1126-6708/2003/05/013} {\bibfield
  {journal} {\bibinfo  {journal} {JHEP}\ }\textbf {\bibinfo {volume} {05}},\
  \bibinfo {pages} {013} (\bibinfo {year} {2003})},\ \Eprint
  {http://arxiv.org/abs/astro-ph/0210603} {arXiv:astro-ph/0210603} \BibitemShut
  {NoStop}%
\bibitem [{\citenamefont {Boruch}\ \emph {et~al.}(2021)\citenamefont {Boruch},
  \citenamefont {Caputa}, \citenamefont {Ge},\ and\ \citenamefont
  {Takayanagi}}]{Boruch:2021hqs}%
  \BibitemOpen
  \bibfield  {author} {\bibinfo {author} {\bibfnamefont {J.}~\bibnamefont
  {Boruch}}, \bibinfo {author} {\bibfnamefont {P.}~\bibnamefont {Caputa}},
  \bibinfo {author} {\bibfnamefont {D.}~\bibnamefont {Ge}}, \ and\ \bibinfo
  {author} {\bibfnamefont {T.}~\bibnamefont {Takayanagi}},\ }\href {\doibase
  10.1007/JHEP07(2021)016} {\bibfield  {journal} {\bibinfo  {journal} {JHEP}\
  }\textbf {\bibinfo {volume} {07}},\ \bibinfo {pages} {016} (\bibinfo {year}
  {2021})},\ \Eprint {http://arxiv.org/abs/2104.00010} {arXiv:2104.00010
  [hep-th]} \BibitemShut {NoStop}%
\bibitem [{\citenamefont {Couvreur}\ \emph {et~al.}(2017)\citenamefont
  {Couvreur}, \citenamefont {Jacobsen},\ and\ \citenamefont
  {Saleur}}]{Couvreur_2017}%
  \BibitemOpen
  \bibfield  {author} {\bibinfo {author} {\bibfnamefont {R.}~\bibnamefont
  {Couvreur}}, \bibinfo {author} {\bibfnamefont {J.~L.}\ \bibnamefont
  {Jacobsen}}, \ and\ \bibinfo {author} {\bibfnamefont {H.}~\bibnamefont
  {Saleur}},\ }\href {\doibase 10.1103/physrevlett.119.040601} {\bibfield
  {journal} {\bibinfo  {journal} {Physical Review Letters}\ }\textbf {\bibinfo
  {volume} {119}} (\bibinfo {year} {2017}),\
  10.1103/physrevlett.119.040601}\BibitemShut {NoStop}%
\bibitem [{\citenamefont {Herviou}\ \emph {et~al.}(2019)\citenamefont
  {Herviou}, \citenamefont {Regnault},\ and\ \citenamefont
  {Bardarson}}]{Herviou_2019}%
  \BibitemOpen
  \bibfield  {author} {\bibinfo {author} {\bibfnamefont {L.}~\bibnamefont
  {Herviou}}, \bibinfo {author} {\bibfnamefont {N.}~\bibnamefont {Regnault}}, \
  and\ \bibinfo {author} {\bibfnamefont {J.~H.}\ \bibnamefont {Bardarson}},\
  }\href {\doibase 10.21468/scipostphys.7.5.069} {\bibfield  {journal}
  {\bibinfo  {journal} {{SciPost} Physics}\ }\textbf {\bibinfo {volume} {7}}
  (\bibinfo {year} {2019}),\ 10.21468/scipostphys.7.5.069}\BibitemShut
  {NoStop}%
\bibitem [{\citenamefont {Chang}\ \emph {et~al.}(2020)\citenamefont {Chang},
  \citenamefont {You}, \citenamefont {Wen},\ and\ \citenamefont
  {Ryu}}]{Chang:2019jcj}%
  \BibitemOpen
  \bibfield  {author} {\bibinfo {author} {\bibfnamefont {P.-Y.}\ \bibnamefont
  {Chang}}, \bibinfo {author} {\bibfnamefont {J.-S.}\ \bibnamefont {You}},
  \bibinfo {author} {\bibfnamefont {X.}~\bibnamefont {Wen}}, \ and\ \bibinfo
  {author} {\bibfnamefont {S.}~\bibnamefont {Ryu}},\ }\href {\doibase
  10.1103/PhysRevResearch.2.033069} {\bibfield  {journal} {\bibinfo  {journal}
  {Phys. Rev. Res.}\ }\textbf {\bibinfo {volume} {2}},\ \bibinfo {pages}
  {033069} (\bibinfo {year} {2020})},\ \Eprint
  {http://arxiv.org/abs/1909.01346} {arXiv:1909.01346 [cond-mat.str-el]}
  \BibitemShut {NoStop}%
\bibitem [{\citenamefont {Holzhey}\ \emph {et~al.}(1994)\citenamefont
  {Holzhey}, \citenamefont {Larsen},\ and\ \citenamefont
  {Wilczek}}]{Holzhey:1994we}%
  \BibitemOpen
  \bibfield  {author} {\bibinfo {author} {\bibfnamefont {C.}~\bibnamefont
  {Holzhey}}, \bibinfo {author} {\bibfnamefont {F.}~\bibnamefont {Larsen}}, \
  and\ \bibinfo {author} {\bibfnamefont {F.}~\bibnamefont {Wilczek}},\ }\href
  {\doibase 10.1016/0550-3213(94)90402-2} {\bibfield  {journal} {\bibinfo
  {journal} {Nucl. Phys. B}\ }\textbf {\bibinfo {volume} {424}},\ \bibinfo
  {pages} {443} (\bibinfo {year} {1994})},\ \Eprint
  {http://arxiv.org/abs/hep-th/9403108} {arXiv:hep-th/9403108} \BibitemShut
  {NoStop}%
\bibitem [{\citenamefont {Calabrese}\ and\ \citenamefont
  {Cardy}(2004)}]{Calabrese:2004eu}%
  \BibitemOpen
  \bibfield  {author} {\bibinfo {author} {\bibfnamefont {P.}~\bibnamefont
  {Calabrese}}\ and\ \bibinfo {author} {\bibfnamefont {J.~L.}\ \bibnamefont
  {Cardy}},\ }\href {\doibase 10.1088/1742-5468/2004/06/P06002} {\bibfield
  {journal} {\bibinfo  {journal} {J. Stat. Mech.}\ }\textbf {\bibinfo {volume}
  {0406}},\ \bibinfo {pages} {P06002} (\bibinfo {year} {2004})},\ \Eprint
  {http://arxiv.org/abs/hep-th/0405152} {arXiv:hep-th/0405152} \BibitemShut
  {NoStop}%
\bibitem [{\citenamefont {Peschel}(2003)}]{Peschel_2003}%
  \BibitemOpen
  \bibfield  {author} {\bibinfo {author} {\bibfnamefont {I.}~\bibnamefont
  {Peschel}},\ }\href {\doibase 10.1088/0305-4470/36/14/101} {\bibfield
  {journal} {\bibinfo  {journal} {Journal of Physics A: Mathematical and
  General}\ }\textbf {\bibinfo {volume} {36}},\ \bibinfo {pages} {L205}
  (\bibinfo {year} {2003})}\BibitemShut {NoStop}%
\bibitem [{\citenamefont {Latorre}\ \emph {et~al.}(2004)\citenamefont
  {Latorre}, \citenamefont {Rico},\ and\ \citenamefont
  {Vidal}}]{Latorre:2003kg}%
  \BibitemOpen
  \bibfield  {author} {\bibinfo {author} {\bibfnamefont {J.~I.}\ \bibnamefont
  {Latorre}}, \bibinfo {author} {\bibfnamefont {E.}~\bibnamefont {Rico}}, \
  and\ \bibinfo {author} {\bibfnamefont {G.}~\bibnamefont {Vidal}},\
  }\href@noop {} {\bibfield  {journal} {\bibinfo  {journal} {Quant. Inf.
  Comput.}\ }\textbf {\bibinfo {volume} {4}},\ \bibinfo {pages} {48} (\bibinfo
  {year} {2004})},\ \Eprint {http://arxiv.org/abs/quant-ph/0304098}
  {arXiv:quant-ph/0304098} \BibitemShut {NoStop}%
\bibitem [{\citenamefont {Casini}\ and\ \citenamefont
  {Huerta}(2009)}]{Casini:2009sr}%
  \BibitemOpen
  \bibfield  {author} {\bibinfo {author} {\bibfnamefont {H.}~\bibnamefont
  {Casini}}\ and\ \bibinfo {author} {\bibfnamefont {M.}~\bibnamefont
  {Huerta}},\ }\href {\doibase 10.1088/1751-8113/42/50/504007} {\bibfield
  {journal} {\bibinfo  {journal} {J. Phys. A}\ }\textbf {\bibinfo {volume}
  {42}},\ \bibinfo {pages} {504007} (\bibinfo {year} {2009})},\ \Eprint
  {http://arxiv.org/abs/0905.2562} {arXiv:0905.2562 [hep-th]} \BibitemShut
  {NoStop}%
\bibitem [{\citenamefont {Mozaffar}\ and\ \citenamefont
  {Mollabashi}(2022)}]{Mozaffar:2021nex}%
  \BibitemOpen
  \bibfield  {author} {\bibinfo {author} {\bibfnamefont {M.~R.~M.}\
  \bibnamefont {Mozaffar}}\ and\ \bibinfo {author} {\bibfnamefont
  {A.}~\bibnamefont {Mollabashi}},\ }\href {\doibase
  10.1103/PhysRevResearch.4.L022010} {\bibfield  {journal} {\bibinfo  {journal}
  {Phys. Rev. Res.}\ }\textbf {\bibinfo {volume} {4}},\ \bibinfo {pages}
  {L022010} (\bibinfo {year} {2022})},\ \Eprint
  {http://arxiv.org/abs/2106.14700} {arXiv:2106.14700 [hep-th]} \BibitemShut
  {NoStop}%
\bibitem [{\citenamefont {Van~Raamsdonk}(2010)}]{VanRaamsdonk:2010pw}%
  \BibitemOpen
  \bibfield  {author} {\bibinfo {author} {\bibfnamefont {M.}~\bibnamefont
  {Van~Raamsdonk}},\ }\href {\doibase 10.1142/S0218271810018529} {\bibfield
  {journal} {\bibinfo  {journal} {Gen. Rel. Grav.}\ }\textbf {\bibinfo {volume}
  {42}},\ \bibinfo {pages} {2323} (\bibinfo {year} {2010})},\ \Eprint
  {http://arxiv.org/abs/1005.3035} {arXiv:1005.3035 [hep-th]} \BibitemShut
  {NoStop}%
\bibitem [{\citenamefont {Swingle}(2012)}]{Swingle:2009bg}%
  \BibitemOpen
  \bibfield  {author} {\bibinfo {author} {\bibfnamefont {B.}~\bibnamefont
  {Swingle}},\ }\href {\doibase 10.1103/PhysRevD.86.065007} {\bibfield
  {journal} {\bibinfo  {journal} {Phys. Rev. D}\ }\textbf {\bibinfo {volume}
  {86}},\ \bibinfo {pages} {065007} (\bibinfo {year} {2012})},\ \Eprint
  {http://arxiv.org/abs/0905.1317} {arXiv:0905.1317 [cond-mat.str-el]}
  \BibitemShut {NoStop}%
\end{thebibliography}%

\appendix

\end{document}